\documentclass[twocolumn,aps,prl,showpacs]{revtex4}
\usepackage{graphicx}
\usepackage{epsf}
\usepackage{amsmath}
\usepackage{amssymb}
\usepackage[usenames]{color}

\begin{document}

\title{Gapless excitations in strongly fluctuating superconducting wires}

\author{{ Dganit Meidan$^1$, Bernd Rosenow$^2$, Yuval Oreg$^3$ and Gil Refael$^4$}\\
{\small \em $^1 $Dahlem Center for Complex Quantum Systems and Institut
f\"{u}r Theoretische Physik, Freie Universit\"{a}t Berlin, 14195
Berlin, Germany\\
$^2$Institut f\"ur Theoretische Physik, Universit\"at Leipzig, D-04103, Leipzig, Germany\\
$^3$Department of Condensed Matter Physics, Weizmann Institute of Science, Rehovot, 76100, ISRAEL\\
$^4$Department of Physics, California Institute of Technology,
Pasadena, California 91125, USA}}

\date{\today}

\begin{abstract}
We study the low temperature tunneling density of states of thin wires where superconductivity is destroyed through quantum phase-slip proliferation. Although this regime is believed to behave as an insulator, we show that for a large temperature range this phase is characterized by a conductivity falling off at most linearly with temperature, and has a gapless excitation spectrum. This novel conducting phase results from electron-electron interaction induced pair breaking. Also, it may help clarify the low temperature metallic features found in films and wires whose bulk realization is superconducting.
\end{abstract}

\maketitle

Phase fluctuations of superconductors are responsible for a broad range of fascinating phenomena. Their effect is particularly dramatic in narrow wires, where the proliferation of phase slips induces a putative superconductor-insulator transition~\cite{SKorshunovJETP1987,SKorshunovEPL1989,PABobbertPRB1990,PABobbertPRB1992,ADZaikinPRL1997,KYuPR2008}. Experiments probing this transition, however, challenge our understanding of the insulating phase, as they exhibit a low-temperature metallic behavior in thin wires where phase-slips proliferate~\cite{ABezryadinNature2000,CNLauRPL2001,ATBollinger2006}. 
Ref.~\cite{ADZaikinPRL1997} discussed the phase-slip induced breakdown of superconductivity in a wire, speculating that a metallic phase arises. Could the strong phase fluctuations induce a finite quasi-particle (QP) density of states that maintains a finite conductivity?

Gapless quasiparticles are well known to exist in superconductors when time reversal symmetry is broken~\cite{AAAbrikosovJETP1960,KMakiPTP1963,PGdeGennesPHY1964,KMakiPTP1964,PFuldePR1966}. Gapless superconductivity, however, also appears due to proximity to metallic contacts where the order parameter is non-uniform~\cite{PFuldePRL1965,PGdeGennesSSC1965}. The pair breaking effect occurs since the relative phase of the two electrons making up a Cooper pair (CP) gets randomized by the perturbation. This begs the question, whether  fluctuating electromagnetic fields due to strong superconducting phase fluctuations and electron-electron interactions, which are manifestly non-uniform and introduce dephasing in normal systems~\cite{AltshulerAronov85}, can lead to the appearance of gapless superconductivity at finite temperatures as well.

In this manuscript we study the low-temperature tunneling density-of-states (tDOS) of the phase-slip proliferated regime. We argue that the scarcity of normal excitations and the blocking of the Cooper pair conduction channel give rise to strong dephasing, through electromagnetic field fluctuations. This, in turn, should lead to pair breaking. From a self-consistent study of the tDOS, we find that at a broad temperature regime (roughly $T<0.1T_c$),  no hard spectral gap exists. Furthermore, because the conductivity is dominated by quasi-particles, it vanishes at most linearly in temperature, as opposed to an exponentially-suppressed conductivity characteristic of a gapped phase. Eventually, at very low temperatures, the QP are localizaed and the metallic phase ceases to be valid. These effects should be manifest in tunneling measurements of wires with an increasing resistance upon cooling.

Our argument follows from the dependence of dephasing on the dissipative response of diffusive electron systems. For this purpose, it is insightful to interpret the response of the wires we consider in terms of coexisting normal quasi-particles and condensed Cooper pairs. In the phase-slip proliferated regime, where the normal resistance of a coherence-length segment obeys $R_{\xi}\gtrsim R_Q=h/4e^2\approx 6.4k\Omega$, the conductivity is dominated by the normal QPs, as long as they remain diffusive. Similarly to normal diffusive systems,  electron-electron interactions lead to the suppression of quantum interference of these diffusive QP after a typical dephasing time $\tau_{\phi} $. Using the fluctuation-dissipation theorem the dephasing is dictated by the 
electrical response of the system, $\sigma(T)$~\cite{BLAltshulerJPC1982}:
\begin{eqnarray}\label{tau_phi}
\tau_\phi(T) 
= \left(\frac{\sigma(T)A}{e^2T\sqrt{2D}}\right)^{2/3}=\tau_\phi^N \left(\frac{\sigma(T)}{\sigma_N}\right)^{2/3},
\label{dephasingtime.eq}
\end{eqnarray}
where $\sigma_N $ is the conductivity in the normal state, $A$ is the cross-section area of the wire, and $D$ is the diffusion constant. When the Cooper-pairs are formed but not condensed due to quantum phase slips proliferation, QP are expected to be scarce, and therefore $\sigma(T)<\sigma_N$. This increases voltage fluctuations and the dephasing rate. 

\begin{figure}[h!]
\begin{center}
\includegraphics[width=0.45\textwidth]{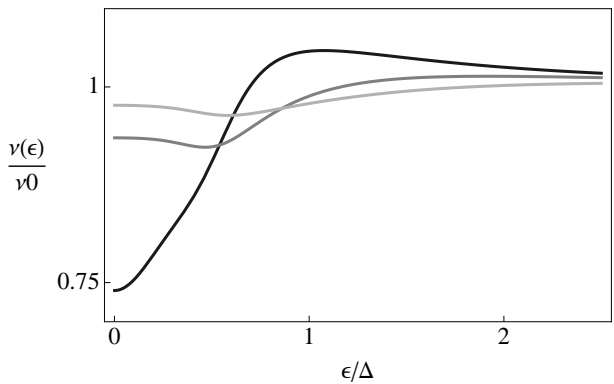}
\caption[0.5\textwidth]{ The tDOS of a fluctuating superconductor given by Eq. \eqref{correction_TDOS}. Here $T/\Delta = 0.1 $, and $\Delta\tau_{KT} = 2 $. Different curves are plotted for different dephasing time with $\Delta \tau_\phi=1,0.5,0.3 $ corresponding to black dark gray and light gray, respectively. The tDOS approaches a constant value for short dephasing times. We note that the initial suppression of the tDOS at $ \epsilon\geq 1/\tau_{KT}$ is a result of the quasi-order of the superconducting phase at times shorter then the correlation time $ 1/\tau_{KT}$.  }
\label{TDOS}
\end{center}
\end{figure}

An enhanced dephasing rate, however, may lead to the breaking of Cooper pairs, and therefore suppress the pairing gap. Indeed, if we assume a hard gap, $\Delta $ in the excitation spectrum we obtain a contradiction. If the QP density is exponentially suppressed $n\sim e^{-\Delta/T} $ due to the pairing gap, then the dephasing rate is exponentially enhanced. A strong dephasing mechanism allows us to consider the effects of pairing on the tDOS perturbatively in the parameter  $\Delta\tau_\phi(T)\ll1$. Such a calculation yields a gapless excitation spectrum that approaches the normal state tDOS (see Fig. \ref{TDOS}  and detailed calculation below), in clear contradiction to the assumption of a finite excitation gap.
This argument is valid at temperatures much lower than the mean-field $T_c$ yet above the localization limit of the QP. Note that thermal superconducting-phase fluctuations may also contribute to dephasing. These contributions are small, however, as QP's dominate transport in this regime.

\begin{figure}[h!]
\begin{center}
\includegraphics[width=0.5\textwidth]{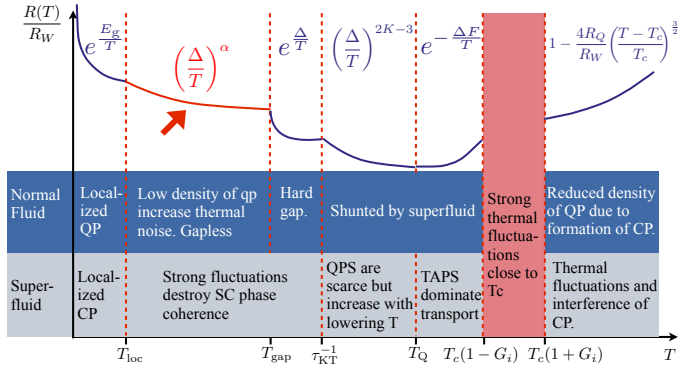}
\caption[0.5\textwidth]{A qualitative phase diagram of a fluctuating wire, whose normal resistance is $ R_W$, as a function of temperature. 
Here $T_c $ is the mean field  transition temperature~\cite{CommentTc} and  $T_Q $ is the temperature at which quantum phase slips (QPS) dominate the response of the system. At $\tau_\textrm{KT}^{-1} $ QPS proliferate, blocking the conductance of the superfluid channel. Gapless excitations appear at $T<T_\textrm{gap} $, where the resistance scales as a power law of the temperature, marked by an arrow. At $T_\textrm{loc} $ QP and CP are localized, and conductance is controlled by their thermally activated hoping  with a typical gap $E_\textrm{g}$. For a summary of all crossover temperature, and typical numerical values see Table I~\ref{table_temp}. For large $R_\xi$ 
the gap regime disappears. Here $K$ is the Luttinger parameter, $\Delta F $ is defined in  \cite{CommentLAMH}, and $G_i $ is the Ginzburg Levanyuk number. The figure is not true to scale. }
\label{phase_diagram}
\end{center}
\end{figure}

Before we discuss the regime of interest, we first summarize the different temperature regimes of a strongly fluctuating  superconducting wire, and their transport signatures. A qualitative phase diagram is depicted in Fig.~\ref{phase_diagram}. 
Above $ T_c$~\cite{CommentTc}, thermal pairing fluctuations and interference effects (Aslamasov Larkin and Maki-Tompson corrections), reduce the resistance.  The formation of incoherent Cooper paris, however, tends to enhance the resistance. Together these reduce the resistance $R(T) = R_W - \frac{h}{e^2}[(T-T_c)/T_c]^{3/2} $, where $ R_W$ is the wires's normal state resistance, and $ (T-T_c)/T_c\ll1$. Near $T_c $ ($|T-T_c|/T_c\ll Gi$ where $Gi=(\frac{7\zeta(3)}{4\pi^2}\frac{R_\xi}{R_Q})^{3/2} $ is the Ginzburg Levanyuk number) strong fluctuations control the resistance. At temperatures below $T_c$, a Cooper pair condensate forms, and shunts the normal excitations. Here thermally activated phase-slippage (TAPS) dominate, and the resistance follows an activation behavior, $ R_{LAMH}(T) = \frac{\pi\hbar^2\Omega}{2e^2k_BT}e^{-\frac{\Delta F}{k_B T}}$ \cite{JSLangerPR1967,DEMcCumberPRB1970,CommentLAMH}. 

At low temperatures $T<T_Q $, where $T_Q $ is defined as $\Delta(T_Q)=T_Q $, quantum phase slips (QPS) dominate the resistance of the superfluid. The behavior of superconducting wires in this regime is dichotomized  by the Luttinger parameter, $ K$, which depends on $R_\xi $, see Eq. \eqref{stiffness} below. In wires with $R_\xi<R_Q $, quantum phase slips are irrelevant and their resistance follows a power law temperature dependence. In this manuscript we focus on wires with  $ R_\xi\gtrsim R_Q$. At intermediate temperatures $1/\tau_{KT}<T<T_Q $, QPS are scarce but their density increases with reducing temperature, resulting in a power law temperature dependence, $R(T)\sim R_W (T/\Delta)^{2 K -3}$. Here $\tau_{KT}^{-1}$ is the typical temperature at which phase slip proliferate, leading to a large phase slip fugacity: $\zeta(1/\tau_{KT}) =1 $. At lower temperature , $T<\tau_{KT}^{-1} $, QPS proliferate and the conductivity is dominated by diffusive quasi-particles.  

In the presence of a hard gap in the excitation spectrum the number of normal quasi particles is exponentially small $n_\textrm{qp}\sim e^{-\Delta/T}$, leading to an exponentially large resistance. However, the following remarkable circumstances may lead to a metallic behavior characterized by a power law resistance. The reduced density of normal excitation, resulting in the formation of Cooper pairs, increases the Nyquist thermal fluctuations of the potential. This finite temperature noise acts as a phase breaker for the superfluid, and consequently may lead to the vanishing of the gap in the excitation spectrum, a situation known as gapless superconductivity. Unlike conventional superconductors, where the superfluid shunts the normal fluid, however,  the proliferation of QPS block the superfluid channel  and the resistance is dominated by these normal quasi-particles giving rise to a metallic behavior. 
The temperature at which the gap vanishes, $ T_\textrm{gap}$,  can be estimated from $\Delta \tau_\phi(T_\textrm{gap})=1 $. Using Eq. \eqref{dephasingtime.eq}, for $ \tau_\phi(T)$ and assuming the conductivity in the hard gap phase follows $\sigma(T)/\sigma_N =e^{-\Delta/T} $,  we find $T_\textrm{gap} \approx \Delta/[\ln(\xi_\textrm{loc}/\xi)+\ln\ln(\xi_\textrm{loc}/\xi)]$.

The gapless regime, $ T<T_\textrm{gap}$, is the main focus of this manuscript. 
In this regime, as we outline below, the resistance of the wire follows a power law  
\begin{eqnarray}\label{power_law}
\sigma(T)/\sigma_N \sim (T/\Delta )^\alpha.
\end{eqnarray} 
Determining the power $ \alpha $ requires summing the perturbation series in $\Delta \tau_\phi $ to infinite order. Nevertheless, from Eq. (\ref{dephasingtime.eq}) ($\sigma(T) \sim \tau_\phi^{3/2} T$) together with the fact that $\tau_{\phi}$ must diverge as $T\rightarrow 0$, we find that the conductivity must follow a sub-linear temperature dependence $\sigma(T)/T\xrightarrow{T\rightarrow 0} \infty $, corresponding to $ \alpha\leq 1$. 
We estimate $ \alpha $ in two different ways: from the leading order correction to the tDOS, and from the leading order correction to the self energy. The former approximation gives $\sigma(T)/\sigma_N =\nu(T)/\nu_0 \approx 1-(\Delta \tau_\phi(T))^2$~\cite{CommentConductivityandTDOS,CommentFiniteTemp}. Substituting the expression for the dephasing time \eqref{dephasingtime.eq}  and solving for the conductivity in the limit $\nu(T)/\nu_0\ll1$, we obtain the power law
$\sigma(T)/\sigma_N \approx \left(\frac{\xi}{\xi_{\textrm{loc}}}\frac{T}{\Delta}\right)$.
Alternatively, the leading order correction to the self energy, which is equivalent to a partial resummation of the infinite series, gives rise to a sub-linear temperature dependence $ \sigma(T)/\sigma_N\sim \left(T/\Delta \right)^{2/5}$. A discussion of this calculation and its validity appears at the end of the manuscript.

The metallic behavior persists as long as the normal fluid remains diffusive. This  breaks down at  low temperatures, where the dephasing length  exceeds the localization length $ L_\phi(T_\textrm{loc})\sim \xi_{loc}$. We estimate $T_\textrm{loc} $ by using the sub linear temperature dependence of the conductivity $ \sigma(T)/\sigma_N\sim \left(T/\Delta \right)^{2/5}$ in Eq. \eqref{dephasingtime.eq}, and find $T_\textrm{loc} = \Delta \left(\xi/\xi_{\textrm{loc}}\right)^{4}$, which for typical wires is well below $T_\textrm{gap} $. Table I~\ref{table_temp} lists the different temperature regimes as well as their numerical values for typical experiments.

\begin{table}\label{table_temp}
\caption{A list of all crossover temperatures, their defining relation, as well as their parametric and numerical values. For the numerical estimates we have used  $\Delta\sim \Delta(T_Q)=T_Q = 0.9 T_c $, $b=1 $, $R_\xi \approx 0.5 R_Q$. In addition, as the localization length satisfies $R_{\xi_\textrm{loc}} = 4 R_Q $, we have $\xi_\textrm{loc}/\xi =8 $. We note that for these choice of parameters, $T_\textrm{gap}\tau_{KT} >1$. }
\begin{tabular}{| c | c | c | c |}
\hline
& Determined from & Parametric  & Numeric \\
\hline
$T_Q $ & $\Delta(T_Q) = T_Q $ &- & $0.9 \,\,T_c$\\
\hline 
$\tau_{KT}^{-1} $ & $ \zeta(\tau_{KT}) =1$  &$\tau_{KT}^{-1} \sim \frac{\Delta}{e^{b|K-K_c|^{-1/2}}}$ & $0.2 \,\,T_c$ \\
\hline
$ T_\textrm{gap}$ & $\Delta \tau_\phi(T_\textrm{gap})=1 $ & $T_\textrm{gap}\sim \frac{\Delta}{\ln{\frac{\xi_\textrm{loc}}{\xi}}+\ln\ln{\frac{\xi_\textrm{loc}}{\xi}}} $& $0.3\,\, T_c$\\
\hline
$ T_\textrm{loc}$ & $L_\phi(T_\textrm{loc})= \xi_{loc}$  &  $ T_\textrm{loc}  \approx \Delta \left(\xi/\xi_{\textrm{loc}}\right)^{4} $ & $0.0002\,\,T_c$\\
\hline
\end{tabular}
\end{table}

The above argument shows that the assumption of a finite dissipation mechanism in the low temperature phase of fluctuating superconducting wires holds self consistently. 
The finite conductivity derived from Eq.~(\ref{dephasingtime.eq}) is  due to diffusive QPs, whose 
interaction dynamics gives rise to a fluctuating potential and hence dephasing. 
The dephasing rate not only gives rise to a finite conductivity as described above, but also causes a finite tDOS. These effects can be  probed  in strongly fluctuating long superconducting wires whose resistance increases as the temperature is reduced, and whose total capacitance is large~\cite{YNazarovPRL1999,DSGolubevPRL2001}. In this regime, 
the energy scale at which the tDOS reaches its maximum value is $1/\tau_\phi $ (see Fig. \ref{TDOS}), which according to these predictions should coincide with the value inferred from an independent measurement of the wire's conductivity following Eq. \eqref{tau_phi}.

Next, we prove the crucial point that if $\tau_\phi $ is indeed small, the QP spectrum cannot be gapped. We later use this calculation to estimate the emerging tDOS in the gapless regime. We carry out a  calculation of the tDOS which is perturbative in  $\Delta\tau_\phi $. In the absence of a pair breaking mechanism, the perturbative correction diverges, marking the opening of a pairing gap.
Conversely, in the presence of strong dephasing, $\Delta \tau_\phi\ll1 $, the QP excitation spectrum may be gapless.
The tDOS is given by
$\nu_\epsilon =-\frac{1}{\pi}\textrm{Im} G^R(r,r,\epsilon)$,
where $G^R(r,r,\epsilon) $ is the retarded Green's function which can be expressed to second order in the pairing amplitude: $G_0 + G_0\Lambda \bar{G}_0\Lambda G_0 \langle\Delta \Delta^\dag\rangle$.
Here $G_0(p,\omega_n)^{-1}= i\omega_n-\xi_p +i/(2\tau) \textrm{Sign}(\omega_n)$, $\bar{G}_0(p,\omega_n)^{-1}= i\omega_n+\xi_p +i/(2\tau) \textrm{Sign}(\omega_n)$  are the disorder averaged free Green's function in the vicinity of the second order phase transition, and $\Lambda$ is the impurity ladder dressed vertex.
We define $\delta\nu(\epsilon) =\frac{\nu(\epsilon)-\nu_0}{\nu_0}=-\frac{1}{\pi}\textrm{Im }I^R(\epsilon)$, where $I^R(\epsilon)=I(i\omega_n\rightarrow\epsilon+i\delta) $ is the analytic continuation of 
\begin{eqnarray}\label{correction_TDOS}
 \!\!\!\! I(\omega_n)=2\pi i\,\text{sign}(\omega_n) T\!\sum_{q,\Omega}\!\! \frac{\Theta(\omega_n(\omega_n+\Omega))\langle\Delta \Delta^\dag\rangle_{q,\Omega}}{(|2\omega_n+\Omega|+Dq^2+\tau_\phi^{-1})^2}.
 \end{eqnarray}

In order to describe correlations of the order parameter in a superconducting wire we examine its microscopic action  obtained from the BCS
Hamiltonian by a Hubbard-Stratanovich transformation followed by
an expansion around the saddle point \cite{DSGolubevPRB2001,DMeidanPhysicaC2008}. 
The low energy excitations of the system are phase fluctuations whose action follow:
\begin{eqnarray}\label{phase_action}
  S[\phi] &=& K/2\int dxdy \left\{(\partial_x\phi)^2+(\partial_y\phi)^2/N_\perp\right\}.
\end{eqnarray}
where $y = v_\rho\tau  $, $N_\perp=p_F^2A /\pi^2 $ and
\begin{eqnarray}\label{stiffness}
  K &=& \frac{4\nu_0 A \Delta_0^2\xi_0^2}{v_\rho}\approx \frac{R_Q}{2R_\xi}.
\end{eqnarray}

The partition function of the superconducting wire whose low energy excitations follow Eq. (\ref{phase_action}) has the same form as a classical partition function of an anisotropic two dimensional XY model. The system described by this model undergoes a Kosterlitz Thouless phase transition between an ordered phase (superconductor) where quantum topological excitations known as phase slip are bound in pairs and a disordered phase where phase slip pairs unbind~\cite{JMKosterlitzJPHYSC1973}. Correlations of the order parameter in the disordered phase decay exponentially:
\begin{eqnarray}\label{corr}
  \langle\Delta(x,\tau)\Delta^{\dag}(0,0) \rangle &=&\Delta_0^2e^{-x/\xi_{KT}}e^{-\tau/\tau_{KT}},
\end{eqnarray}
over a typical length  $\xi_{KT} $, and time $\tau_{KT} $. This gives
\begin{eqnarray}\label{phase_correlator}
  \langle\Delta\Delta^{\dag}\rangle_{q,\Omega}&=&\frac{\Delta_0^2\xi_{KT}\tau_{KT}}{(1+q^2\xi_{KT}^2)(1+\Omega^2\tau_{KT}^2)}.
\end{eqnarray}

Using Eqs. \eqref{correction_TDOS} and \eqref{phase_correlator} we calculate the corrections to the tDOS in a fluctuating superconductor.
The result is shown in Fig.~\ref{TDOS}. We have assumed a proliferation of phase slips which requires to be in the regime $ T \lesssim 1/\tau_{KT} \lesssim \Delta$. The perturbative correction to the tDOS is small if $\Delta \lesssim 1/\tau_\phi $, which in general is not satisfied in experiments. However, for illustrative purposes and to stay in the regime where our approximation is justified, Fig.~\ref{TDOS} shows the tDOS for $\Delta \tau_\phi=1,0.5,0.3 $, and the extension to the regime $1/\tau_\phi < \Delta$ is discussed below. 
In Fig.~\ref{TDOS} one sees that the zero energy density of states decreases with increasing $\Delta \tau_\phi$. As $\tau_\phi$ grows with 
decreasing temperature, we expect that $\nu(\epsilon)$ develops a pseudogap, consistent with the 
power law temperature dependence of $\sigma$.

One might question the consistency of our calculation  in the limit $\Delta \tau_\phi\ll 1 $, because the gapless tDOS we find can act as a shunt resistor, suppress quantum phase fluctuations, and restore local phase coherence~\cite{SKorshunovJETP1987,SKorshunovEPL1989,PABobbertPRB1990,GRefaelPRB2007}, contrary to the assumed strongly fluctuating regime. 
This discrepancy is resolved since the phase-fluctuating regime \eqref{phase_correlator} 
corresponds to a gapped kernel for the pairing field~\cite{CommentGappedKernel}. Hence, while gapless Fermionic excitations may introduce
a dissipation term of the form $R_s |\Omega||\Delta|^2$  in the action for the
pairing field, the substantial mass term in the action dictates the long
time correlations rendering the dissipation unimportant, and the system
remains in the strongly fluctuating phase. Indeed, in the small $\Omega$
limit, the pairing-field action~\cite{CommentGappedKernel} with an $R_s>0$ term coincides with
the Hertz-Millis action for the metallic phase of a strongly fluctuating
superconductor~\cite{SSachdevPRL2004,AVLopatinPRL2005,ADMaestroPRB2008}. In addition,
the phase diagram of dissipatively shunted Josephson junction chains
also exhibits a disordered phase which is insensitive to small
dissipation~\cite{SKorshunovJETP1987,SKorshunovEPL1989,PABobbertPRB1990,GRefaelPRB2007}.

Let us now describe how to estimate the emerging tDOS and consequently the conductivity in the gapless regime, beyond the perturbative limit.
Note that while the perturbative correction presented above is no longer small  for a long dephasing time $\Delta \tau_\phi>1 $, it does not diverge and hence a calculation of the full 
density of states correction is possible. Similarly to the  case of pure Coulomb interaction,  $\tau_\phi$ is in principle due to higher order corrections in the perturbation series in the interaction strength. 
As such a calculation is beyond the scope of this letter, we use the dephasing rate obtained from the conductivity according to Eq.~(\ref{dephasingtime.eq}) in Eq.~(\ref{correction_TDOS}). Although it cannot be trusted, calculating the leading order to the self energy  may give additional qualitative information about the tDOS in the regime $\Delta \tau_\phi\gg1 $, see details in the supplementray material. 
 This leads to a finite sub-gap density of states at low but finite temperatures, which vanishes at $ T\rightarrow 0$. In the limit $T\tau_{KT},T\tau_\phi\ll1 $ we find $\sigma(T)/\sigma_0 \approx \nu(T)/\nu_0\sim 1/(\Delta \tau_\phi(T)) = (\xi/\xi_\textrm{loc} T/\Delta )^{2/5} $, were the last equality was obtained using \eqref{dephasingtime.eq}~\cite{CommentConductivityandTDOS,CommentFiniteTemp}. Similar models considering the role of a fluctuating pair correlator in a ballistic system in two dimensions revealed a QP peak in the spectral function~\cite{TSenthilPRB2009, TMicklitzPRB2009}.

In conclusion, we studied 
the tDOS of a superconducting wire in the QPS proliferated regime.  We found that in the disordered phase, associated with QPS proliferation, the conductivity has a sub-linear temperature dependence, and is dominated by quasi-particles with a substantial  sub-gap excitation spectrum. This novel metallic phase may be related to the metallic behavior observed in low dimensional films and wires whose bulk realization is a superconductor. These predictions can be experimentally tested by tunneling measurements. In this regime, the energy scale at which the tDOS is expected to reduce below its value in the normal phase, $ \nu_0$ is $1/\tau_\phi $, which according to these predictions should coincide with the value inferred from an independent measurement of the wire's conductivity following Eq. \eqref{tau_phi}. 

We acknowledge helpful discussions with E. A. Demler, B. I. Halperin, P.A. Lee, T. Senthil. 
We thank the Packard Foundation, BSF, DIP, the Alexander von Humboldt Foundation
and the Aspen Center for Physics.


\begin{thebibliography}{10}

\bibitem{SKorshunovJETP1987}
S.~Korshunov,
\newblock {S}ov. Phys. JETP {\bf 66}, 872 (1987).

\bibitem{SKorshunovEPL1989}
S.~Korshunov,
\newblock Europhys. Lett. {\bf 9}, 107 (1989).

\bibitem{PABobbertPRB1990}
P.~A. Bobbert, R.~Fazio, G.~Sch\"{o}n, and G.~T. Zimanyi,
\newblock Phys. Rev. B {\bf 41}, 4009 (1990).

\bibitem{PABobbertPRB1992}
P.~A. Bobbert, R.~Fazio, G.~Sch\"on, and A.~D. Zaikin,
\newblock Phys. Rev. B {\bf 45}, 2294 (1992).

\bibitem{ADZaikinPRL1997}
A.~D. Zaikin, D.~S. Golubev, A.~van Otterlo, and G.~T. Zimanyi,
\newblock Phys. Rev. Lett. {\bf 78}, 1552 (1997).

\bibitem{KYuPR2008}
K.~Y. Arutyunov, D.~S. Golubev, and A.~D. Zaikin,
\newblock Physics Reports {\bf 464}, 1 (2008).

\bibitem{ABezryadinNature2000}
A.~Bezryadin, C.~N. Lau, and M.~Tinkham,
\newblock Nature {\bf 404}, 971 (2000).

\bibitem{CNLauRPL2001}
C.~N. Lau, N.~Markovic, M.~Bockrath, A.~Bezryadin, and M.~Tinkham,
\newblock Phys. Rev. Lett. {\bf 87}, 217003 (2001).

\bibitem{ATBollinger2006}
A.~T. Bollinger, A.~Rogachev, and A.~Bezryadin,
\newblock Europhys. Lett. {\bf 76}, 505 (2006).

\bibitem{AAAbrikosovJETP1960}
A.~A. Abrikosov and L.~P. Gorkov,
\newblock Zh. Eksp. Teor. Fiz {\bf 39}, 1781 (1960).

\bibitem{KMakiPTP1963}
K.~Maki,
\newblock Progr. Theoret. Phys. {\bf 29}, 333 (1963).

\bibitem{PGdeGennesPHY1964}
P.~G. de~Gennes and M.~Tinkham,
\newblock Physics {\bf 1}, 107 (1964).

\bibitem{KMakiPTP1964}
K.~Maki,
\newblock Progr. Theoret. Phys. {\bf 32}, 29 (1964).

\bibitem{PFuldePR1966}
P.~Fulde and K.~Maki,
\newblock Phys. Rev. {\bf 141}, 275 (1966).

\bibitem{PFuldePRL1965}
P.~Fulde and K.~Maki,
\newblock Phys. Rev. Lett. {\bf 15}, 675 (1965).

\bibitem{PGdeGennesSSC1965}
P.~G. de~Gennes and S.~Mauro,
\newblock Solid State Commun. {\bf 3}, 381 (1965).

\bibitem{AltshulerAronov85}
B.~L. Altshuler and A.~G. Aronov,
\newblock {\em Electron-Electron Interaction in Disordered Conductors}, Modern
  Problems in Condensed Matter Sciences Vol.~10 (North-Holland, Amsterdam,
  1985), .

\bibitem{BLAltshulerJPC1982}
B.~L. Altshuler, A.~G. Aronov, and D.~E. Khmelnitsky,
\newblock J. Phys. C {\bf 15}, 7367 (1982).

\bibitem{CommentTc}
We will assume henceforth that corrections due to Coulomb repulsion in a
  diffusive system, which reduce $T_c $, are taken into account.

\bibitem{CommentLAMH}
Here $\Delta F= 8\sqrt{2}/3 (H_c^2/8\pi)A\xi\approx 0.83 (R_Q/R_\xi)k_B
  T_c(1-T/T_c)^{3/2}$, $\Omega = L/\xi (\Delta F/k_B T)^{1/2} \tau_{GL}^{-1} $,
  and $\tau_{GL}=\pi \hbar/(8k_B (T_c-T)) $.

\bibitem{JSLangerPR1967}
J.~S. Langer and V.~Ambegaokar,
\newblock Phys. Rev. {\bf 164}, 498 (1967).

\bibitem{DEMcCumberPRB1970}
D.~E. McCumber and B.~I. Halperin,
\newblock Phys. Rev. B {\bf 1}, 1054 (1970).

\bibitem{CommentConductivityandTDOS}
In the weakly interacting limit considered throughout this paper, we assume
  $\sigma(T)\propto \nu(T) $.

\bibitem{CommentFiniteTemp}
At finite temperatures, the normal part of the conductivity is given by the
  thermal average $\sigma_N(T)\propto \int d\epsilon \left(-\partial
  f(\epsilon,T)/\partial \epsilon\right)\nu(\epsilon)$. In the limit
  $T\tau_{KT}\ll1$ and $T\ll\Delta\ll \tau_\phi^{-1}$, we can replace
  $\nu(\epsilon)\approx\nu(0) $.

\bibitem{YNazarovPRL1999}
Y.~Nazarov,
\newblock Phys. Rev. Lett. {\bf 82}, 1245 (1999).

\bibitem{DSGolubevPRL2001}
D.~S. Golubev and A.~D. Zaikin,
\newblock Phys. Rev. Lett. {\bf 86}, 4887 (2001).

\bibitem{DSGolubevPRB2001}
D.~S. Golubev and A.~D. Zaikin,
\newblock Phys. Rev. B {\bf 64}, 014504 (2001).

\bibitem{DMeidanPhysicaC2008}
D.~Meidan, Y.~Oreg, G.~Refael, and R.~A. Smith,
\newblock Physica C {\bf 468}, 341 (2008).

\bibitem{JMKosterlitzJPHYSC1973}
J.~M. Kosterlitz and D.~J. Thouless,
\newblock J. Phys. C {\bf 6}, 1181 (1973).

\bibitem{GRefaelPRB2007}
G.~Refael, E.~Demler, Y.~Oreg, and D.~S. Fisher,
\newblock Phys. Rev. B {\bf 75}, 014522 (2007).

\bibitem{CommentGappedKernel}
The action for $ \Delta$ in this regime is $S[\Delta]\approx
  v_\phi/\Delta^2\int \frac{d\Omega}{2\pi} \frac{dq}{2\pi}\left(\xi_{KT}^{-2} +
  q^2+\Omega^2/v_\phi^2\right)|\Delta_{q,\Omega}|^2$.

\bibitem{SSachdevPRL2004}
S.~Sachdev, P.~Werner, and M.~Troyer,
\newblock Phys. Rev. Lett. {\bf 92}, 237003 (2004).

\bibitem{AVLopatinPRL2005}
A.~V. Lopatin, N.~Shah, and V.~M. Vinokur,
\newblock Phys. Rev. Lett. {\bf 94}, 037003 (2005).

\bibitem{ADMaestroPRB2008}
A.~D. Maestro, B.~Rosenow, N.~Shah, and S.~Sachdev,
\newblock Phys. Rev. B {\bf 77}, 180501 (2008).

\bibitem{TSenthilPRB2009}
T.~Senthil and P.~A. Lee,
\newblock Phys. Rev. B {\bf 79}, 245116 (2009).

\bibitem{TMicklitzPRB2009}
T.~Micklitz and M.~R. Norman,
\newblock Phys. Rev. B {\bf 80}, 220513 (2009).

\end{thebibliography}
\end{document}


\title{Supplementary Material for\\ ``Gapless excitations in strongly fluctuating superconducting wires''}

\author{{ Dganit Meidan$^1$, Bernd Rosenow$^2$, Yuval Oreg$^3$ and Gil Refael$^4$}\\
{\small \em $^1 $Dahlem Center for Complex Quantum Systems and Institut
f\"{u}r Theoretische Physik, Freie Universit\"{a}t Berlin, 14195
Berlin, Germany\\
$^2$Institut f\"ur Theoretische Physik, Universit\"at Leipzig, D-04103, Leipzig, Germany\\
$^3$Department of Condensed Matter Physics, Weizmann Institute of Science, Rehovot, 76100, ISRAEL\\
$^4$Department of Physics, California Institute of Technology,
Pasadena, California 91125, USA}}

\date{\today}
\maketitle

\section{Microscopic phase action}
In order to describe correlations of the order parameter in a superconducting wire we examine its microscopic action  obtained from the BCS
Hamiltonian by a Hubbard-Stratanovich transformation followed by
an expansion around the saddle point \cite{DSGolubevPRB2001,DMeidanPhysicaC2008}. 
In the low temperature limit, this
yields \cite{DSGolubevPRB2001,DMeidanPhysicaC2008}:
\begin{eqnarray}\label{effective_action}
\nonumber
  S &=& \nu_0A\Delta_0^2\int_0^L dx\int_0^{1/T} d\tau\left\{\frac{\rho^2}{2}\left[\ln\left(\rho^2\right)-1\right] +2\xi_0^2\rho^2\left[\phi'^2+\frac{1}{v_\phi^2}\dot{\phi}^2 \right]+\xi_0^2\left[\rho'^2+\frac{1}{v_{\rho}^2}\dot{\rho}^2
  \right]\right\},
\end{eqnarray}
where $L$ and $A$ are the wire's length and cross section,
respectively, $\xi_0^2=\pi D/8\Delta_0 $,
$v_\rho=\sqrt{(3\pi/2)D\Delta_0}$ the amplitude velocity,
$v_\phi=\sqrt{\pi D\Delta_0(2AV_c\nu_0+1)}\propto v_\rho\sqrt{N_\perp} $ the phase velocity,
$V_c$ the Fourier transform of the short range Coulomb
interaction, $N_\perp=p_F^2A /\pi^2$ is the number of one dimensional channels in the wire, $\nu_0$ the density of states, $D$ the electronic
diffusion constant in the normal state, and the SC order parameter
is parameterized as $\Delta = \Delta_0\rho e^{i\phi}$, with
$\Delta_0$, the mean field solution. Rescaling the imaginary time by $y = v_\rho\tau  $, the low energy excitations of the system are phase fluctuations whose action follow:
\begin{eqnarray}\label{phase_action}
  S[\phi] &=& K/2\int dxdy \left\{(\partial_x\phi)^2+(\partial_y\phi)^2/N_\perp\right\}.
\end{eqnarray}
where the phase stiffness is
\begin{eqnarray}\label{stiffness}
  K &=& \frac{4\nu_0 A \Delta_0^2\xi_0^2}{v_\rho}\approx \frac{R_Q}{2R_\xi}.
\end{eqnarray}
The system described by this model undergoes a Kosterlitz Thouless phase transition between an ordered phase (superconductor) and a disordered phase where phase slip pairs unbind~\cite{JMKosterlitzJPHYSC1973}. Correlations of the order parameter in the disordered phase decay exponentially:
\begin{eqnarray}\label{corr}
  \langle\Delta(x,\tau)\Delta^{\dag}(0,0) \rangle &=&\Delta_0^2e^{-x/\xi_{KT}}e^{-\tau/\tau_{KT}},
\end{eqnarray}
over a typical length  $\xi_{KT} $, and time $\tau_{KT} $. This corresponds to
\begin{eqnarray}\label{phase_correlator}
  \langle\Delta\Delta^{\dag}\rangle_{q,\Omega}&=&\frac{\Delta_0^2\xi_{KT}\tau_{KT}}{(1+q^2\xi_{KT}^2)(1+\Omega^2\tau_{KT}^2)}.
\end{eqnarray}

\section{Leading order correction to the tunneling density of states of a fluctuating superconductor}

The tDOS is given by
\begin{eqnarray}
\nu_\epsilon =-\frac{1}{\pi}\textrm{Im} G^R(r,r,\epsilon)=-\frac{1}{\pi}\textrm{Im} \int \frac{d^3p}{(2\pi)^3}G^R(p,\epsilon),
\end{eqnarray}
where $G^R(r,r,\epsilon) $ is the retarded Green's function which can be expressed to second order in the pairing amplitude:
\begin{widetext}
\begin{eqnarray}\label{PT}
  G(p,\omega_n) 
 &=& G_0(p,\omega_n)+T\sum_{q,\Omega}G_0(p,\omega_n)\Lambda(q,\omega_n,\omega_n+\Omega) G_0(p+q,\omega_n+\Omega)\Lambda(q,\omega_n+\Omega,\omega_n)G_0(p,\omega_n) \langle\Delta \Delta^\dag\rangle_{q,\Omega} .
\end{eqnarray}
\end{widetext}
Here:
\begin{eqnarray}
\nonumber
G_0(k+q,\omega)^{-1} &=& i(\omega)+\frac{i}{2\tau}\textrm{sign}(\omega) -\xi\\
\nonumber
 \Lambda(\omega,\omega+\Omega,q) &=& \frac{1}{2\tau} \frac{\Theta(\omega(\omega+\Omega))}{|2\omega+\Omega|+Dq^2 +1/\tau_\phi},
\end{eqnarray}
and correlations of the order parameter are given by Eq. \eqref{phase_correlator}.
The density of states is then given by $\delta\nu(\epsilon) =\frac{\nu(\epsilon)-\nu_0}{\nu_0}=-\frac{1}{\pi}\textrm{Im }I^R(\epsilon)$, where $I^R(\epsilon)=I(i\omega_n\rightarrow\epsilon+i\delta) $ is the analytic continuation of 
\begin{eqnarray}\label{correction_TDOS}
I(\omega_n)=2\pi i\,\text{sign}(\omega_n) T\!\sum_{q,\Omega}\!\! \frac{\Theta(\omega_n(\omega_n+\Omega))\langle\Delta \Delta^\dag\rangle_{q,\Omega}}{(|2\omega_n+\Omega|+Dq^2+\tau_\phi^{-1})^2}.
 \end{eqnarray}
 Using Eq. \eqref{phase_correlator} to describe the phase fluctuations in a phase-slip proliferated wire, in the low energy limit $\tau_\phi \ll\tau_{KT}$ we may approximate Eq. \eqref{correction_TDOS} as
\begin{widetext}
\begin{eqnarray}\label{apprx_TDOS}
\nonumber
 I(\omega_n)&\approx&2\pi i\,\text{sign}(\omega_n)\frac{\Delta_0^2}{(|2\omega_n|+\tau_\phi^{-1})^2}T\!\sum_{\Omega}\frac{\Theta(\omega_n(\omega_n+\Omega))\tau_{KT}}{1+\Omega^2\tau_{KT}^2}\int\frac{dq}{2\pi}\frac{\xi_{KT}}{1+q^2\xi_{KT}^2}\\
 &=&\frac{\pi i\,\text{sign}(\omega_n)\Delta_0^2}{(|2\omega_n|+\tau_\phi^{-1})^2}\left\{\frac{i}{4\pi}\left[\Psi\left(\frac{1}{2}+\frac{\omega_n}{2\pi T}+\frac{i}{2\pi T\tau_{KT}}\right)-\Psi\left(\frac{1}{2}+\frac{\omega_n}{2\pi T}-\frac{i}{2\pi T\tau_{KT}}\right)\right]+\frac{1}{2}\coth \frac{1}{2T\tau_{KT}}\right\},
 \end{eqnarray}
 \end{widetext}
 where $\Psi (z)$ is the digamma function.
 
\section{Leading order correction to the self energy}
The leading order correction to the self energy is given by $G^{-1} = G_0^{-1}-\Sigma $ with:
\begin{eqnarray}
\Sigma = \sum_qT\sum_\Omega \bar{G}(k+q,\omega+\Omega) \langle \Delta \Delta^\dag\rangle_{q,\Omega} \Lambda^2(\omega,\omega+\Omega,q).
\end{eqnarray}
The integral over fermionic momentum is dominated  by $\xi\approx 1/\tau $. Since, $\omega\tau,\Omega\tau,Dq^2\tau \ll1 $, we can approximate $\bar{G}(k+q,\omega+\Omega)\approx \bar{G}(k,\omega) $.
This gives
\begin{eqnarray}
\nonumber
\Sigma& \approx& \bar{G}(k,\omega)\sum_qT\sum_\Omega \frac{\Theta(\omega(\omega+\Omega))}{4\tau^2(|2\omega+\Omega|+Dq^2 +1/\tau_\phi)^2}\langle \Delta \Delta^\dag\rangle_{q,\Omega} \\
&\equiv& \bar{G}(k,\omega) A(\omega).
\end{eqnarray}
Using this expression for the self energy we can write the green's function as:
\begin{eqnarray}
G(k,\omega)^{-1} &=& i(\omega)+\frac{i}{2\tau}\textrm{sign}(\omega) -\xi_{k}-\Sigma(\omega)\\
&=&i\tilde{\omega}-\xi - \frac{1}{i\tilde{\omega}+\xi}A(\omega),
\end{eqnarray}
where $\tilde{\omega}  = \omega+\frac{1}{2\tau}\textrm{sign}(\omega)$. 
The density of states is given by:
\begin{eqnarray}
\nonumber
\nu(i\omega) &=&-\frac{1}{\pi}\int dk G(k,\omega) 
=\frac{1}{\pi} \nu_0\int d\xi \frac{i\tilde{\omega}+\xi}{\tilde{\omega}^2+\xi^2+ A(\omega)}\\
&=& \nu_0 \frac{i\tilde{\omega}}{\sqrt{\tilde{\omega}^2+ A(\omega)}}
\end{eqnarray}
where the odd integral over $\xi $ vanishes. In the limit of $\omega\tau \ll1 $ we have:
\begin{eqnarray}
\nu(i\omega) &=&   \nu_0 \frac{i\textrm{sign}(\omega)}{\sqrt{1+ 4\tau^2A(\omega)}}\approx   \nu_0 \frac{i\textrm{sign}(\omega)}{\sqrt{ 4\tau^2A(\omega)}}
\end{eqnarray}
where the last approximation is valid beyond the perturbative limit where $4\tau^2A(\omega) \gg1 $.

In order to evaluate $4\tau^2A(\omega) $, we note that $4\tau^2A(\omega)  = \frac{I(\omega)}{2\pi i \textrm{sign}(\omega)}$ where $I(\omega) $ is given by Eq. \eqref{correction_TDOS}. Using Eq.~\eqref{apprx_TDOS} in the limit $T\tau_{KT} \ll1 $ we find: 
\begin{eqnarray}
\nonumber
4\tau^2A(\omega)  &=& \frac{\Delta_0^2}{2}\frac{1}{(|2\omega|+1/\tau_\phi)^2}\left\{\frac{i}{4\pi}\left[\Psi(\frac{1}{2} +\frac{\omega}{2\pi T}+\frac{i}{2\pi T\tau_{KT}})-\Psi(\frac{1}{2} +\frac{\omega}{2\pi T}-\frac{i}{2\pi T\tau_{KT}})\right]+\frac{1}{2}\coth (\frac{1}{2T\tau_{KT}})\right\}\\
&=&\frac{\Delta_0^2}{2}\frac{1}{(|2\omega|+1/\tau_\phi)^2} \left\{ \frac{i}{4\pi}\left[ i\pi+2i \pi T\tau_{KT} -2 i \omega \tau_{KT}\right]+1/2  \right\}
\end{eqnarray}
Here we have assumed $\omega \sim T \ll 1/\tau_{KT} $. Performing the analytic continuation $i\omega \rightarrow \epsilon +i\delta $  we find
\begin{eqnarray}
\nonumber
4\tau^2A(i\omega \rightarrow \epsilon +i\delta ) 
\nonumber
&=&\frac{\Delta_0^2}{2}\frac{1}{(-2i\epsilon +1/\tau_\phi)^2} \left\{ \frac{i}{4\pi}\left[ i\pi+2i \pi T\tau_{KT} -2 \epsilon \tau_{KT}\right]+1/2  \right\}\\
\nonumber
&=&\frac{\Delta_0^2}{2}\frac{1}{(-2i\epsilon +1/\tau_\phi)^2} \left(\frac{1}{4}-\frac{i}{2\pi} \epsilon \tau_{KT}-\frac{1}{2}T\tau_{KT}\right)\\
\end{eqnarray}
The density of states is given by
\begin{eqnarray}
\nu(\epsilon ) &=& \Im \nu(i\omega\rightarrow \epsilon+i\delta) = \Im \left[ i  \nu_0\frac{2}{\Delta_0}  \frac{(-2i\epsilon +1/\tau_\phi) }{\sqrt{1/2-\frac{i}{\pi} \epsilon \tau_{KT}-T\tau_{KT}}}\right]. 
\end{eqnarray}
In the low temperature limit $T\tau_{KT},T\tau_\phi \ll1 $, we can replace $\nu(T)=- \int d\epsilon \nu(\epsilon) \frac{df}{d\epsilon}\approx \nu(\epsilon =0,T)$, leading to:
\begin{eqnarray}
\frac{\nu(T )}{\nu_0} &=& \frac{2\sqrt{2}}{\Delta_0\tau_\phi(T)}.
\end{eqnarray}